\documentclass[cits]{PoS}

\title{The high-energy emission from GX~339--4 as seen with INTEGRAL and XMM-Newton}

\ShortTitle{The high-energy emission from GX~339--4}

\author{\speaker{M.~D.~Caballero-Garcia}\thanks{This work is based on observations
made with {\it INTEGRAL}, an ESA science mission with
instruments and science data centre funded by ESA member states and
with the participation of Russia and the USA and on observations with {\it XMM-Newton},
an ESA science mission with instruments and contributions directly funded by ESA
member states and the USA (NASA).}\\
        University of Cambridge, Institute of Astronomy, Cambridge CB3 0HA, UK\\
        E-mail: \email{mcaballe@ast.cam.ac.uk}}

\author{J.~M.~Miller\\
        Department of Astronomy, University of Michigan, 500 Church Street, Ann Arbor, USA, MI 48109\\
        E-mail: \email{jonmm@umich.edu}}

\author{A.~C.~Fabian\\
        University of Cambridge, Institute of Astronomy, Cambridge CB3 0HA, UK\\
        E-mail: \email{acf@ast.cam.ac.uk}}

\author{, on behalf of a larger collaboration team. }

\abstract{GX~339--4 is a well-known microquasar. In this contribution we show the 
obtained results with the INTEGRAL and XMM-Newton observatories of the outburst 
undertaken on 2007. The observations cover spectral evolution from the hard, soft 
intermediate states to the high/soft state. Spectral hardening correlated with the 
appearance of an skewed Fe line is detected during one of the observations during 
the soft intermediate state. In all spectral states joint XMM/EPIC-pn, JEM-X, ISGRI 
and SPI data were fit with the hybrid thermal/non-thermal Comptonization model 
(EQPAIR). With this model a non-thermal component seems to be required by the data 
in all the observations. Our results imply evolution in the coronal properties, 
the most important one being the transition from a compact corona in the first 
observation to the disappearance of coronal material in the second and re-appearance 
in the third. We discuss the results obtained in the context of possible physical 
scenarios for the origin and geometry of the corona and its relation to black hole states.}

\FullConference{7th INTEGRAL Workshop\\
		 September 8-11 2008\\
		 Copenhagen, Denmark}

\begin{document}

\section{Black Hole States}

When a Black Hole (hereafter BH) transient starts an ouburst, it evolves through different states 
(low/hard, hard intermediate, soft intermediate, high/soft states) characterized by different 
spectral, timing, optical, IR and radio properties. For a recent prescription of the state 
classification scheme we refer to \cite{homanbell05} and \cite{belloni05}. In the earlier 
times of this science (before 1990s) it was thought that the state evolution of a black hole 
was driven by $\dot{M}$ (\cite{esin97}). But, since some states can span a large variation in
luminosities, it was suggested (\cite{homan01}) that other parameter may play a role in these 
state transitions. This parameter could be, e.g. the coronal compactness of the 
high-energy emission.

\begin{figure}
\centering
\includegraphics[width=1.\textwidth]{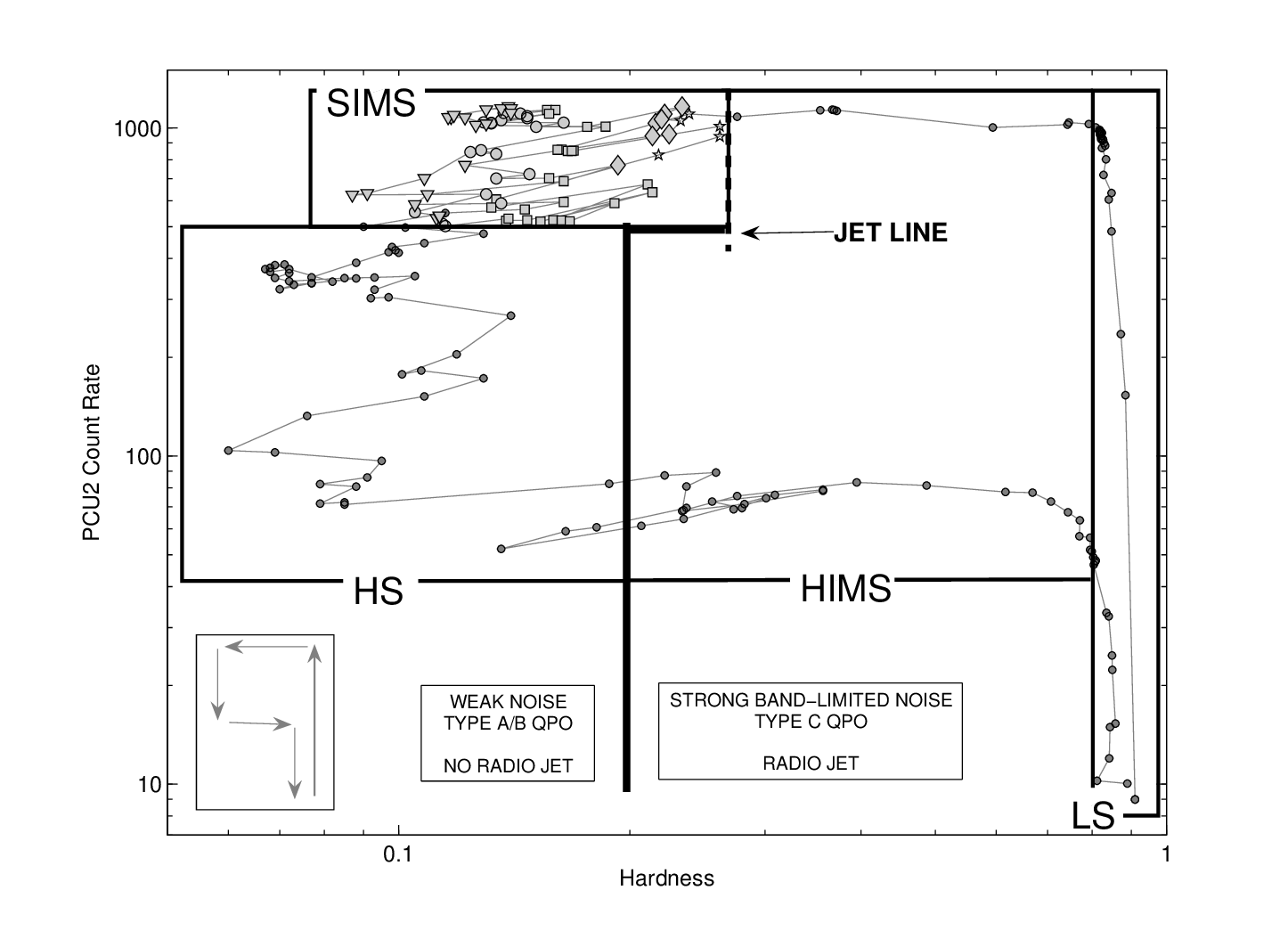}
\caption{Hardness-Intensity diagram of the 2002/2003 outburst of GX~339--4 as observed by the
RXTE PCA. The gray lines mark the state transitions described in the text. The inset on the lower left 
shows the general time evolution of the outburst along the {\bf q}-shaped pattern. From \cite{belloni05}.}
\label{fig1}
\end{figure}

In the state evolution of most Black Hole Binaries (BHB) (\cite{belloni05}, \cite{homanbell05}), these
start their outbursts in the Low/Hard (LH) state and evolve to the High/Soft (HS) state, passing through Hard InterMediate 
(HIMS) and Soft InterMediate (SIMS; formerly called Very/High) states and, afterwards, they return to the LH state
showing a noticeable decrease in flux of the order of 50\%, in a hysteresis behaviour
(\cite{miyamoto95}, \cite{smith02} and \cite{maccarone03}; see evolution in Figure \ref{fig1}). 
In the following, we list some spectral characteristics of all the different 
states from the recent prescription, apart from the 
quiescent state in which black hole transients spend most part of their lifes.

\begin{itemize}

\item{Low/hard state (LS): this is the state in which the outbursts begin and end. The X-ray spectrum is characterized by very low disk
emission and very important high-energy emission in the form of a powerlaw with photon index in the range ${\Gamma}=1.3-1.4$. A
high-energy cut-off is usually seen (\cite{sunyaev80} and \cite{grove98}), associated to the kinetic temperature of the thermal
distribution of electrons in the Comptonizing corona. Sometimes, low frequency QPOs are observed. Flat-spectrum radio emission is
observed, associated to a compact jet ejection (\cite{fender04}).}

\item{Intermediate (soft/hard) states (SIMS/HIMS): showing both bright disk and high-energy powerlaw emission components. Photon index
is within the range ${\Gamma}=1.5-2.5$. The few instances of HFQPOs appeared in the SIMS (formerly called as Very/High state). Just
before the transition to the SIMS, \cite{fender04} suggested that the jet velocity increased rapidly, giving rise to a fast relativistic
jet.}

\item{High/Soft state (HS): the disk component is the dominant in the spectrum, with a weak powerlaw high-energy emission. No core
radio emission is observed (\cite{fender04}). Some timing properties (\cite{wijnands99}), that were thought to be characteristic of the LH and
HIMS are still present in this state, although in a much weaker form. This would suggest a common origin in the characteristics
found in these states. No high-energy cut-off is observed in this state (\cite{grove98}). }

\end{itemize}

\section{Source of the high-energy emission}

The X/${\gamma}$-ray spectra (1--1\,000\,keV) from black holes can be described by a soft disk emission ($<10$\,keV) plus
a high-energy emission ($>20$\,keV) in the form of a powerlaw. The soft X-ray spectra often shows signatures from the inner parts of the
accretion disk, such as the relativistic Fe line (6.4--6.97\,keV) and the correspondent reflection bump (20--30\,keV), both being different
aspects of the same physical origin (i.e. fluorescence of the Fe ions and Compton back-scattering, respectively, both being the most obvious
reactions of an irradiated disk by a high-energy source; \cite{george91}).

The high-energy emission comes from the (inverse) Comptonization of the soft seed photons from the inner accretion disk by a corona
(consituted by electrons and positrons). Both the
geometry and location of this corona are a matter of debate. In the former model (\cite{esin97}), this corona was
filling the inner regions of a truncated disk. \cite{markoff03}, \cite{markoff05} showed that the base of a jet could replace
an extended corona for the high-energy emission source. Independetly, \cite{miller06} found the existence of an inner disk in GX~339--4
during LH state from X-ray observations. All these issues can be better explained
if the base of a jet is the source of hard X-ray emission. This scenario is very tempting, due to the non-thermal emission already
observed in BHs during different states (\cite{joinet07} in the LH, \cite{malzac06} and \cite{gierlinski99} in the intermediate and
HS states, respectively), which could be understood as, e.g. synchrotron processes occurring at the base of the jet.

We will show in the following that the base of a jet, which evolves in properties (opacity, compacticity and distance from the BH) during 
the outburst, could give rise to the X/${\gamma}$ properties already observed in the black hole candidate GX~339--4.

\section{{\it INTEGRAL} and {\it XMM-Newton} observations of GX~339--4}

In \cite{caballero08} we present simultaneous {\it XMM-Newton} and {\it INTEGRAL} observations of the luminous black
hole transient and relativistic jet source GX~339--4. GX~339--4 started an outburst on November
of 2006 and our observations were undertaken from January to March of 2007 (see Table \ref{tab1}). We triggered five
{\it INTEGRAL} and three {\it XMM-Newton} target of Opportunity observations within this period.
Our data cover different spectral states, namely Hard Intermediate (obs. 1), Soft Intermediate (obs. 2 and 3) and
High/Soft (obs. 4 and 5).

\begin{table}
\centering
\begin{tabular}{llll}
   Epoch & XMM-Newton ID & INTEGRAL & XMM-Newton \\ 
\hline
           &           & (yyyy/mm/dd)  & (UTC hh:mm ; yyyy/mm/dd) \\ 
           &   &       &           \\ 
\hline
           &   &       &           \\ 
         1 &  $-$       & 2007/01/30-02/01 & $-$               \\ 
           &   &       &           \\ 
         2 & 0410581201 & 2007/02/17-19    & 00:03--04:44 ; 2007/02/19 \\ 
           &   &       &           \\ 
         3 & 0410581301 & 2007/03/04-06    & 11:15--11:15 ; 2007/03/05 \\ 
           &   &       &           \\ 
         4 &   $-$      & 2007/03/16-18    &   $-$          \\ 
           &   &       &           \\ 
         5 & 0410581701 & 2007/03/29-31    & 14:34--20:07 ; 2007/03/30 \\ 
           &   &       &           \\ 
\end{tabular}
\caption{{\it INTEGRAL} and {\it XMM-Newton} Observations LOG.}
\label{tab1}
\end{table}

\begin{table}
\centering
\begin{tabular}{llllll}
\hline\noalign{\smallskip}
Obs. number & ${\ell}_{h}/{\ell}_{s}$ & ${\ell}_{nth}/{\ell}_{h}$ & ${\tau}_{p}$ & $kT_{e}$\,(keV) & $[{\Omega}/2{\pi}]$ \\
\noalign{\smallskip}\hline\noalign{\smallskip}
1 & $3.9^{+0.6}_{-0.2}$ & $0.40_{-0.03}^{+0.15}$ & $2.39_{-0.18}^{+0.15}$ & $27.5{\pm}1.2$ & $0.38^{+0.06}_{-0.04}$ \\
2 & $0.05_{-0.01}^{+0.003}$ & $0.90{\pm}0.10$ & $<0.02$ & $69{\pm}4$ & $0.40_{-0.04}^{+0.3}$ \\
3 & $0.28^{+0.03}_{-0.01}$ & $0.84{\pm}0.03$ & $1.41_{-0.06}^{+0.03}$ & $10.8{\pm}0.3$ &  1(f)\\
4 & $0.24^{+0.02}_{-0.005}$ & $0.49^{+0.02}_{-0.01}$ & $2.5{\pm}0.5$ & $4.3{\pm}0.8$ &  1(f) \\
5 & $0.13^{+0.01}_{-0.02}$ & $0.38{\pm}0.02$ & $0.89^{+0.04}_{-0.05}$ & $10.5{\pm}0.7$ & $0.72^{+0.16}_{-0.10}$ \\
\noalign{\smallskip}\hline
\end{tabular}
\caption{Best-fit parameters of the joint XMM/EPIC-pn, JEM-X, ISGRI and SPI spectra for the 5 obs. Fits have been performed simultaneously with EQPAIR combined with LAOR.}
\label{tab:1}       
\end{table}

The hybrid thermal/non-thermal Comptonization EQPAIR model (\cite{coppi99}) provides
the injection of a non-thermal electron distribution
with Lorentz factors between ${\Gamma}_{min}$ and ${\Gamma}_{max}$ and
a powerlaw spectral index ${\Gamma}_{inj}$. The cloud is illuminated
by soft thermal photons emitted by an accretion disk. These photons
serve as seed for inverse Compton scattering by both thermal and non-thermal
electrons. The system is characterized by the power (i.e. luminosity)
$L_{i}$ supplied by its different components. We express each of them
dimensionlessly as a compactness parameter,
${\ell}_{i}=L_{i}{\sigma}_{\rm T}/(R m_{e}c^{3})$,
where $R$ is the characteristic dimension and ${\sigma}_{\rm T}$
the Thompson cross-section of the plasma. Thus, ${\ell}_{s}$, ${\ell}_{th}$,
${\ell}_{nth}$ and ${\ell}_{h}={\ell}_{th}+{\ell}_{nth}$ correspond to
the power in a soft disk entering the plasma, thermal electron heating,
electron acceleration and the total power supplied to the plasma.
The total number of electrons (not including ${\rm e}^{+}{\rm e}^{-}$ pairs)
is determined by ${\tau}_{\rm T}$,
the corresponding Thompson optical depth, measured from the center
to the surface of the scattering region. If we consider injection from pairs
${\rm e}^{+}{\rm e}^{-}$, then the total optical depth of the thermalized
scattering electrons/pairs is expected to be ${\tau}_{\rm T}{\geq}{\tau}_{\rm P}$.
We used the LAOR model (\cite{laor91}) to model the relativistic iron line emission, with the
emissivity index (${\beta}$) free and tied to the opposite value of that of the EQPAIR.

In the EQPAIR model, emission of the disk/corona system is modeled by a spherical hot plasma cloud with
continuous acceleration of electrons illuminated by soft seed photons from the accretion disk. At high-energies
the distribution of electrons is non-thermal, but at low energies a Maxwellian distribution with temperature
$kT_{e}$ is established.

In general, the spectral shape is insensitive to the exact value of the compactness, but
it depens strongly on the compactness ratios (${\ell}_{h}/{\ell}_{s}$ and  ${\ell}_{nth}/{\ell}_{h}$).
Thus, we froze ${\ell}_{s}$ to a fiducial value (${\ell}_{s}=10$), as commonly
reported for other sources with similar characteristics (e.g. \cite{ibragimov05}). In the
fits reported below, we fit the data with a powerlaw distributed injection of electrons and compare
with the results obtained with a mono-energetic distributed injection of them as well. The former
distribution is expected in the case of shock acceleration of particles, while the second
could be achieved in reconnection events that are expected to power the corona (\cite{galeev79}).

\begin{figure}
\centering
\includegraphics[width=1.\textwidth]{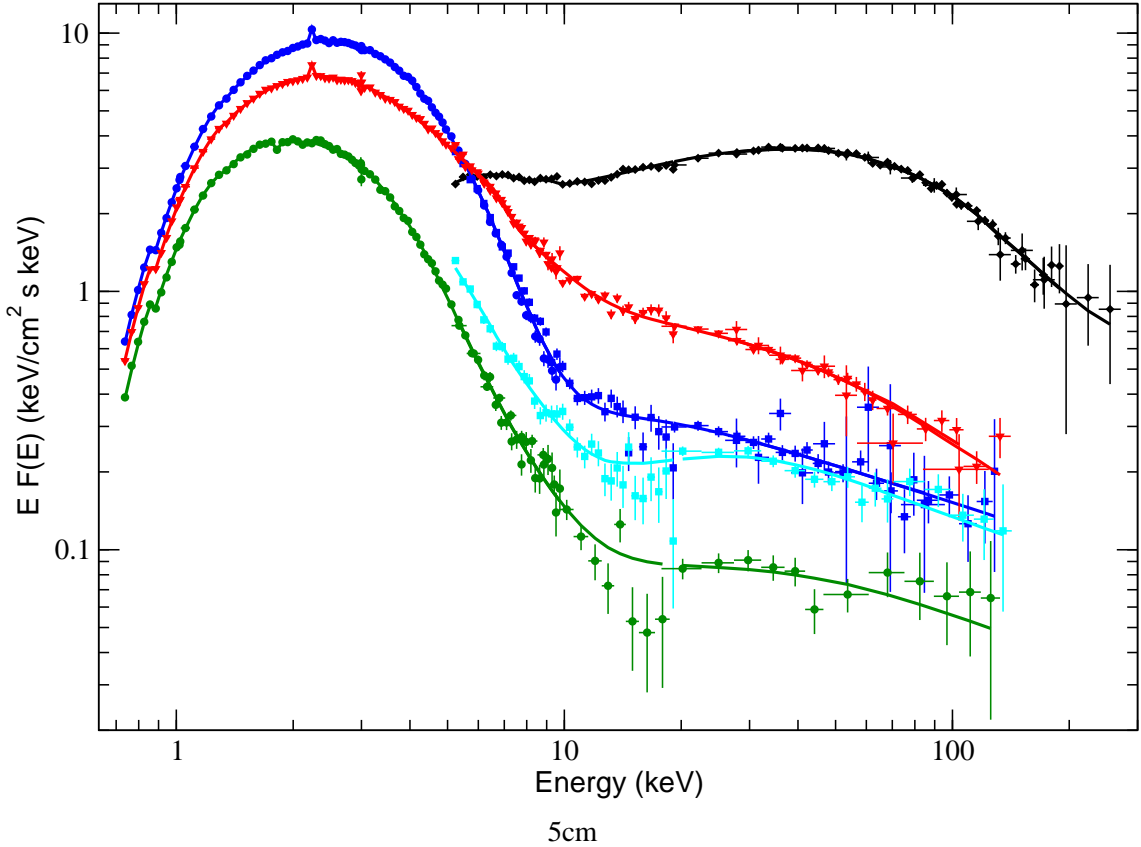}
\vbox{5cm}
\caption{Unfolded spectra from epoch 1 to 5 (black, blue, red, cyan and green, respectively).}
\label{spec_state}
\end{figure}

Spectra and unfolded models of the five different periods have been plotted in Figure \ref{spec_state}.
In Table \ref{tab:1} we show the evolution of the most important parameters inferred from the model.

The results obtained by applying EQPAIR fits to the data indicate a high value for the coronal compactness for obs. 1,
but within the range of values found in the literature. For obs. 3, 4 and 5 this value is high as well (when compared
to that obtained in obs. 2). We thus confirm the correlation between coronal compactness and covering fraction of the cold
reflecting material by \cite{nowak02} for obs. 2 to 5. The high value of the coronal compactness found for obs. 1 (HIMS)
would indicate that the Comptonizing high-energy source is compact in size. This would be in agreement with the proposed
scenario of \cite{markoff05}, in which the base of the jet could be the source of the Comptonizing elecrons.
The fact that we are detecting the thermal cut-off would be consistent with the detection of the coronal emission as well.
Otherwise, for obs. 2 (SIMS), the values for both the coronal compactness and opacity found
are extremely low. Moreover, the kinetic temperature found for the thermal electron distribution of the corona
is very high (and close to the high-energy limit indeed). We understand them as issues indicative of the lack of
coronal emission during this observation. During obs. 3 to 5 (SIMS to HS), both the coronal compactness and opacity
increase again (accompanied with the significant detection of a relativistic line in obs. 3), thus indicating re-appearance of
the corona after obs. 2.

We fit the data with a powerlaw distributed injection of accelerated non-thermal electrons 
(with Lorentz factors in the range of ${\Gamma}=1.3-100$) and compared
with the results obtained with a mono-energetic distributed injection (with Lorentz factor ${\Gamma}=5$) of them as well. 
The former distribution is expected in the case of shock acceleration of particles, while the second
could be achieved in reconnection events that are expected to power the corona (\cite{galeev79}). We found that the first model
is better than the second for epochs 1, 3, 4 and 5 (applying an F-test and looking at the residuals, i.e.
see Table \ref{tab2} and Figure \ref{spec_both}). This is indicative of a corona of particles distributed at different
speeds being the source of the high-energy emission. However, for epoch 2, albeit the first proposed scenario is not
discarded, and contrary to the remainder epochs, a mono-energetic distribution of particles is a good description
of the spectrum. It seems that magnetic reconnection events are driving the high-energy emission in this epoch.

\begin{table}
\centering
\begin{tabular}{llll}
   Epoch & Mono-energetic acceleration  & Powerlaw acceleration & F-test probability\\
\hline
           & (red. chi-square; number of d.o.f.)  & (red. chi-square; number of d.o.f.) &   \\
           &       &          &    \\
\hline
           &       &          &    \\
         1 & 1.1 (83)   &   1.0 (82)    &  0.003  \\
           &       &          &    \\
         2 & 1.7 (143)  &   1.7 (142)   &  $-$        \\
           &       &          &    \\
         3 & 1.9 (139)  &   1.5 (138)   &  7.1e-09 \\
           &       &          &    \\
         4 & 4.5 (45)   &   1.5 (44)    & 2.8e-12 \\
           &       &          &    \\
         5 & 1.7 (100)  &   1.6 (99)    & 0.008  \\
           &       &          &    \\
\end{tabular}
\caption{Statistics obtained with EQPAIR fits to the data with both mono-energetic and powerlaw acceleration cases.}
\label{tab2}
\end{table}

\begin{figure}
\centering
\includegraphics[angle=270.,width=.4\textwidth]{powerlaw_acc1.eps}
\includegraphics[angle=270.,width=.4\textwidth]{powerlaw_mono1.eps}
\includegraphics[angle=270.,width=.4\textwidth]{powerlaw_acc2.eps}
\includegraphics[angle=270.,width=.4\textwidth]{powerlaw_mono2.eps}
\includegraphics[angle=270.,width=.4\textwidth]{powerlaw_acc3.eps}
\includegraphics[angle=270.,width=.4\textwidth]{powerlaw_mono3.eps}
\includegraphics[angle=270.,width=.4\textwidth]{powerlaw_acc4.eps}
\includegraphics[angle=270.,width=.4\textwidth]{powerlaw_mono4.eps}
\includegraphics[angle=270.,width=.4\textwidth]{powerlaw_acc5.eps}
\includegraphics[angle=270.,width=.4\textwidth]{powerlaw_mono5.eps}
\vbox{5cm}
\caption{Unfolded spectra from epoch 1 to 5 applying EQPAIR fits to the data with both mono-energetic and powerlaw acceleration cases. The vertical jumps between the different spectra are due to the different cross-calibration factors between the instruments (not accounted for in this unfolded representation).}
\label{spec_both}
\end{figure}

\begin{figure}
\centering
\includegraphics[width=1.\textwidth]{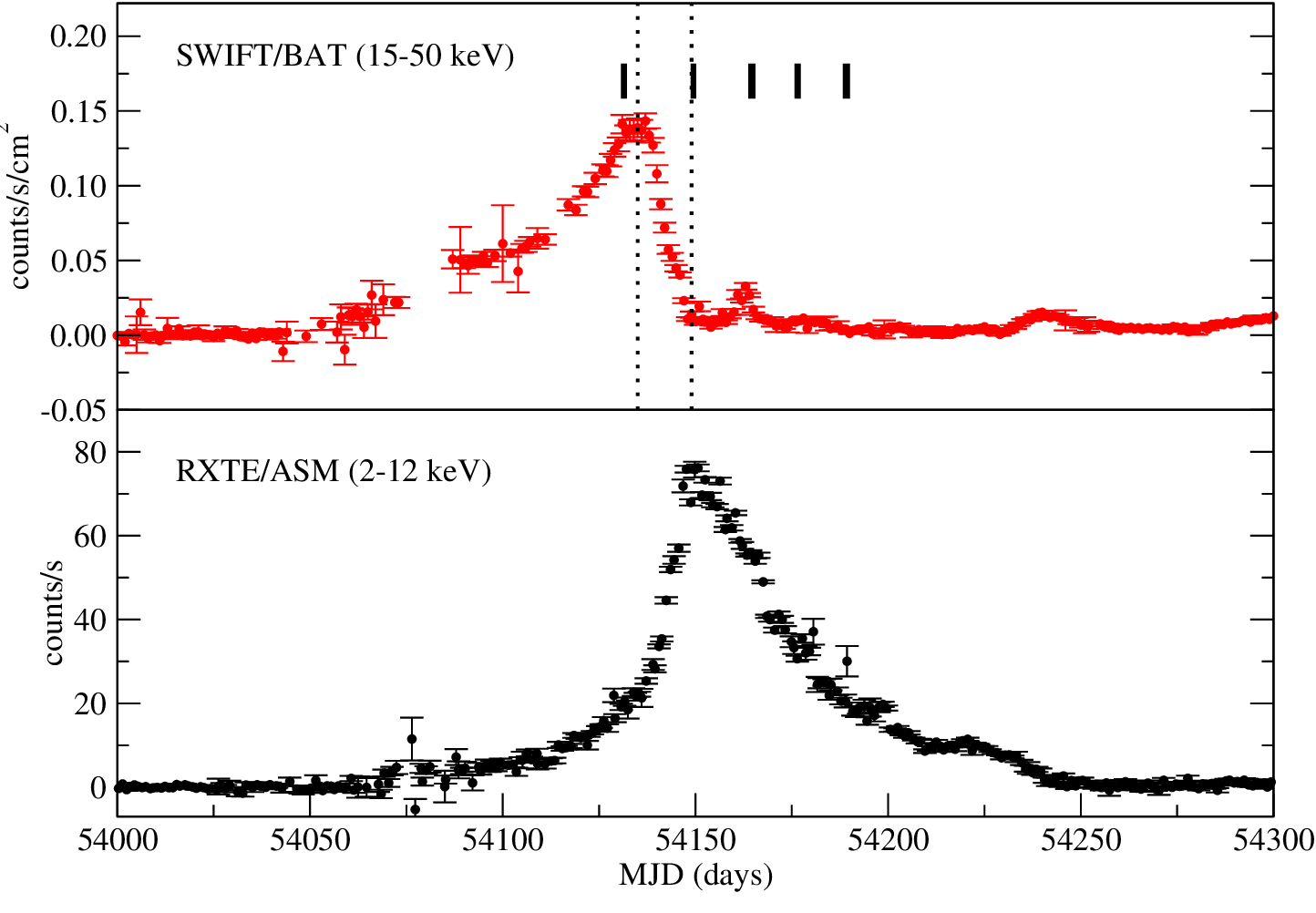}
\vbox{8cm}
\caption{SWIFT/BAT and RXTE/ASM daily light curves of GX~339--4 during the overall outburst in 2007 (red and black dotted lines, respectively), illustrating
the spectral evolution between the different states. Intervals of time in which the INTEGRAL observations
were undertaken (solid black lines) and the period of time when the radio ejection events were detected (within both dotted black vertical lines) are also shown. }
\label{lcurves_swift}
\end{figure}

Thus, we conclude that we detect spectral evolution in our data compatible with disappearance of a part \footnote{Since the model still requires a high-energy component.}
of the corona in epoch 2 (SIMS). This was followed by its re-appearance in epoch 3 (SIMS)
and maintained in epochs 4 and 5. Giving strength to this
interpretation is the fact that \cite{corbel07} detected a series of plasma ejection events during 4--18 of
February (2007) in radio, just previously to our observations of epoch 2. Also, the sudden increase in flux in the 15--50\,keV
of the SWIFT/BAT light curve (Figure \ref{lcurves_swift}) of $15\%$ could be related to changes occurring in the source of the
high-energy emission during the transition from epochs 2 to 3.
The possible disappearance of the corona during epoch 2 resembles what claimed by
\cite{rodriguez08} in the case of GRS~1915+105. This behavior could be understood
as the fact that the ejected medium is the coronal material responsible for the hard X-ray emission.
The spectra of epochs 1, 3, 4 and 5 show a significant fraction of non-thermal particles as well, indicating that
it could be due to other processes apart from thermal Comptonization. For example, synchrotron or self-synchrotron
emission (\cite{markoff03},\cite{markoff05}) occurring at the base of a jet.

\end{document}